\documentclass[floatfix,twocolumn,showpacs,aps,tightenlines]{revtex4-1}
\usepackage{graphicx}
\usepackage{color}
\usepackage{psfrag}
\usepackage{dcolumn}% Align table columns on decimal point
\usepackage{bm}% bold math

\usepackage{amssymb}
\usepackage{amsmath}
\usepackage{amsfonts}
\usepackage{longtable}
\usepackage{xspace}
\usepackage{mathrsfs}
\usepackage[export]{adjustbox}

\newcommand{\beq}{\begin{equation}}
\newcommand{\eeq}{\end{equation}}

\newcommand{\be}{\begin{equation}}
\newcommand{\ee}{\end{equation}}
\newcommand{\bea}{\begin{eqnarray}}
\newcommand{\eea}{\end{eqnarray}}
\newcommand{\bes}{\begin{subequations}}
\newcommand{\ees}{\end{subequations}}

\newcommand{\scri}{\mathscr{I}}

\usepackage{bm}

\begin{document}

\title{The Fourth RIT binary black hole simulations catalog: Extension to Eccentric Orbits}
\author{James Healy}
\author{Carlos O. Lousto}
\affiliation{Center for Computational Relativity and Gravitation,
School of Mathematical Sciences,
Rochester Institute of Technology, 85 Lomb Memorial Drive, Rochester,
New York 14623}

\date{\today}

\begin{abstract}
This fourth release of the RIT public
catalog of numerical relativity black-hole-binary
waveforms \url{http://ccrg.rit.edu/~RITCatalog}
consists of 1881 accurate simulations that include 446 precessing
and 611 nonprecessing quasicircular/inspiraling binary
systems with mass ratios $q=m_1/m_2$ in the
range $1/128\leq q\leq1$ and individual spins up to $s/m^2=0.95$;
and 824 in eccentric orbits in the range $0<e\leq1$.
The catalog also provides
initial parameters of the binary, 
trajectory information, peak radiation, and final remnant black hole
properties. 
The waveforms are corrected for the center of mass drifting
and are extrapolated to future null infinity.
As an application of this waveform catalog we reanalyze all of the
peak radiation and remnant properties to find new, simple,
correlations among them, valid in the presence of eccentricity,
for practical astrophysical usage.
\end{abstract}

 \maketitle

\section{Introduction}\label{sec:Intro}

Since the 
breakthroughs~\cite{Pretorius:2005gq,Campanelli:2005dd,Baker:2005vv} in
numerical relativity solved the binary black hole problem
those techniques have been used to explore the 
dynamics of spinning black-hole binaries beyond the post-Newtonian
regime.
The first generic, long-term precessing
black-hole binary evolutions (i.e., without any symmetry) were
performed in Ref.~\cite{Campanelli:2008nk}, where a detailed
comparison with post-Newtonian $\ell=2,3$ waveforms modes was made.
Numerical simulations have then explored the corners of
parameter space, including near extremal spins $\chi=S_i/m_i^2=0.99$,
binaries \cite{Lovelace:2014twa,Zlochower:2017bbg},
mass ratios as small as 
%$q=1/100$ in Refs.~\cite{Lousto:2010ut,Sperhake:2011ik},
$q=m_1/m_2=1/128$ in Ref.~\cite{Lousto:2020tnb},
and large initial separations, $R=100m$, in Ref.~\cite{Lousto:2013oza}.
Numerical relativity has also proved to be able to produce
very long waveforms starting at proper separations of $25m$
for a precessing binary in \cite{Lousto:2015uwa}
and for a nonspinning binary with 176 orbits
to merger in Ref.~\cite{Szilagyi:2015rwa}.
%\NOTE{ADD HERE HEC?}.

Other important studies include the exploration of the {\it hangup}
effect, i.e. the role individual black-hole spins play to delay or
accelerate their merger \cite{Campanelli:2006uy, Hannam:2007wf,Hemberger:2013hsa,Healy:2018swt}, the determination of the magnitude
and direction of the {\it recoil} velocity of the final merged black
hole \cite{Campanelli:2007ew,Campanelli:2007cga,Herrmann:2007ex,Pollney:2007ss,Baker:2006vn,Gonzalez:2007hi, Schnittman:2007ij,Lousto:2011kp},
and the {\it flip-flop} of individual spins during the orbital phase
\cite{Lousto:2014ida, Lousto:2015uwa, Lousto:2016nlp},
the {\it L-flip} of the orbital angular momentum leading to beaconing
of gravitational waves \cite{Lousto:2018dgd}, as well as generic
precession dynamics \cite{Schmidt:2010it,Lousto:2013vpa,Pekowsky:2013ska,Ossokine:2015vda} and the inclusion
of those dynamical effects to construct surrogate models for gravitational
waveforms \cite{Blackman:2017dfb,Blackman:2017pcm,Varma:2018mmi}.

Numerical relativity
predictions of the gravitational waveforms from the late inspiral,
plunge, merger, and ringdown of black-hole-binary systems (BHB)
based on the breakthroughs~\cite{Pretorius:2005gq,Campanelli:2005dd,Baker:2005vv} helped to accurately identify the first direct
detection \cite{TheLIGOScientific:2016wfe} of gravitational waves with
that of binary black hole systems \cite{Abbott:2016blz,Abbott:2016nmj,TheLIGOScientific:2016pea,Abbott:2016wiq} and match them to 
targeted supercomputer simulations \cite{Abbott:2016apu,TheLIGOScientific:2016uux,Lovelace:2016uwp,Healy:2017xwx}.
There have been several significant efforts to collect numerical
relativity simulations into waveform catalogs released by the
SXS collaboration~\cite{Mroue:2013xna,Blackman:2015pia, Chu:2015kft,Boyle:2019kee},
Georgia Tech.~\cite{Jani:2016wkt}, and RIT~\cite{Healy:2017psd,Healy:2019jyf,Healy:2020vre}.
The third RIT catalog release \cite{Healy:2020vre}
reaching 777 simulations, was recently used for parameter estimation of all BBH
O1/O2 gravitational waves signals from LIGO-Virgo in \cite{Healy:2020jjs} and to assess
theoretical definitions of the center of mass evolution of BBH systems in
\cite{Tassone:2021blw}, among several other applications.

In this paper we describe a new release of the public waveform catalog
by the RIT numerical relativity group that total 1881
simulations by adding a new set of 824 eccentric orbits waveforms,
and 134 nonprecessing and 146 precessing quasicircular inspiraling binaries .
The catalog includes all waveform modes $\ell\leq4 $ modes of
the Weyl scalar $\psi_4$ and the strain $H$
(both extrapolated to null-infinity) and is updated to correct for
the center of mass displacement during inspiral and after merger.
The catalog can be accessed from the site \url{http://ccrg.rit.edu/~RITCatalog}.

This paper is organized as follows.  In Section \ref{sec:FN} we
briefly summarize the methods and criteria for producing the numerical
simulations.  In Sec.~\ref{sec:Catalog}
we describe the relevant BHB parameters, the file format, and the
content of the data in the catalog. In Sec.~\ref{sec:eccentricity}
we describe the production of 824 eccentric simulations. 
In Sec.~\ref{sec:correlations} we seek simple correlations among
the black hole merger remnant and peak waveform parameters.
Sec.~\ref{sec:Discussion} concludes with a discussion of the future use of this
catalog 
for parameter inference of new gravitational waves events and
the extensions of this work to longer, more generic precessing binaries.

\section{Full Numerical Evolutions}\label{sec:FN}

The simulations in the RIT Catalog were evolved using the {\sc
LazEv} code~\cite{Zlochower:2005bj} implementation of the moving puncture
approach~\cite{Campanelli:2005dd}. In most cases we use the BSSNOK
formalism of evolutions systems~\cite{Nakamura87, Shibata95, Baumgarte99}.
(except for the very highly spin holes, $\chi>0.9$ where we use CCZ4~\cite{Alic:2011gg}).
For the runs in the catalog, metadata, such as finite-difference
orders, Kreiss-Oliger dissipation orders, and Courant factors~\cite{Lousto:2007rj,Zlochower:2012fk,Healy:2016lce} are included as references associated
with each run (where detailed studies have been performed). 

The {\sc LazEv} code uses the {\sc Cactus}~\cite{cactus_web} /
{\sc Carpet}~\cite{Schnetter-etal-03b} / 
{\sc EinsteinToolkit}~\cite{Loffler:2011ay, einsteintoolkit} 
infrastructure.  The {\sc Carpet} mesh refinement driver provides a
``moving boxes'' style of mesh refinement.
We use {\sc AHFinderDirect}~\cite{Thornburg2003:AH-finding} to locate
apparent horizons.  We first measure the magnitude of the horizon spin 
$S_H$, using the {\it isolated horizon} (IH) algorithm 
as  implemented in Ref.~\cite{Campanelli:2006fy}.
We can then calculate the horizon
mass via the Christodoulou formula 
${m_H} = \sqrt{m_{\rm irr}^2 + S_H^2/(4 m_{\rm irr}^2)}\,,$
where $m_{\rm irr} = \sqrt{A_H/(16 \pi)}$ and $A_H$ is the surface area
of the horizon. 

To compute the numerical (Bowen-York) initial data, we use
the {\sc TwoPunctures}~\cite{Ansorg:2004ds} code.  To
determine quasi-circular orbits  we use the third post-Newtonian
order techniques described in~\cite{Healy:2017zqj}.
To produce eccentric orbital parameters, we reduce the tangential quasicircular
linear momentum by a factor $(1-\epsilon)$.
We evaluate eccentricity during evolution via the simple formula,
as a function of the separation of the holes, $d$,
$e_d=d^2\ddot{d}/m$, as given in \cite{Campanelli:2008nk}.

In Ref.~\cite{Ruchlin:2014zva} to generate more realistic initial data
with reduced spurious gravitational wave content, we have chosen a
background ansatz of conformal superposition of (boosted) Kerr spatial
metrics.  These new initial data, denoted as HiSpID, are relevant for
nonspinning as well as very highly spinning black holes in a
binary~\cite{Zlochower:2017bbg}, and high energy
collisions~\cite{Healy:2015mla}. To generate those data we generalized
the {\sc TwoPunctures} code~\cite{Ansorg:2004ds} to solve a coupled
system of the Hamiltonian and momentum constraints. We use these data
for evolving highly spinning binaries with intrinsic spins
$\chi_i=S_i/m_i^2>0.9$.

We measure radiated energy, linear momentum, and angular momentum, in
terms of the radiative Weyl scalar $\psi_4$, using the formulas
provided in Refs.~\cite{Campanelli:1998jv, Lousto:2007mh}.  As
described in Ref.~\cite{Nakano:2015pta}, we use the Teukolsky equation
to analytically extrapolate expressions for $r \psi_4$ to $\scri^+$

In Ref. \cite{Healy:2020vre} we describe a practical implementation of
the center of mass correction to account for a linear shift in time of
the origin of coordinates for the multipole decomposition of the
waveform (See Ref. \cite{Boyle:2015nqa}).  This shift can be performed
a posteriori, by de-mixing modes at each time step from a center of
coordinates that is linearly moving \cite{Kelly:2012nd}.

Several cases presented in this catalog have been studied in detail to
evaluate typical errors.  In Appendix A of Ref.~\cite{Healy:2014yta},
in Appendix B of Ref.~\cite{Healy:2016lce}, and in
Ref.~\cite{Healy:2017mvh}, we performed convergence studies for
different mass ratios and spins of the binaries.  In in
Ref.~\cite{Lovelace:2016uwp} and Ref.~\cite{Healy:2017abq} we compared
the RIT waveforms with those produced completely independently by the
SXS collaboration finding excellent agreement, convergence towards
each others results and matching of individual modes up to $l=5$.

We conclude that the waveforms at the resolutions provided in this
catalog are well into the convergence regime (roughly converging at
4th-order with resolution), that the horizon evaluated quantities such
as the remnant final mass and spins have errors of the order of 0.1\%,
and that the radiatively computed quantities such as the recoil
velocities and peak luminosities are evaluated at a typical error of
5\%.

A more detailed account of our full numerical techniques is described in
Sect. II of Ref.~\cite{Healy:2020vre} and references therein.

\section{The Catalog}\label{sec:Catalog}

%\begin{widetext}
\begin{figure*}
  \includegraphics[angle=0,width=0.494\columnwidth]{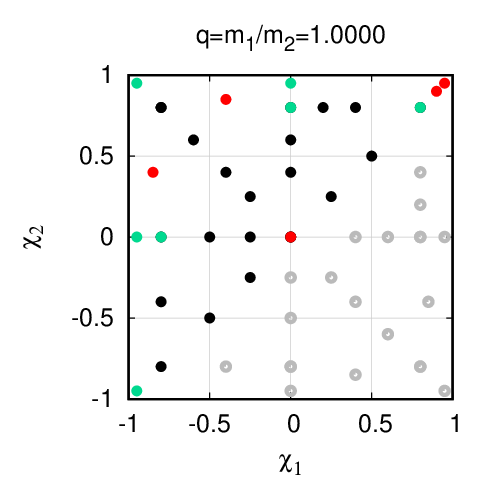}
  \includegraphics[angle=0,width=0.494\columnwidth]{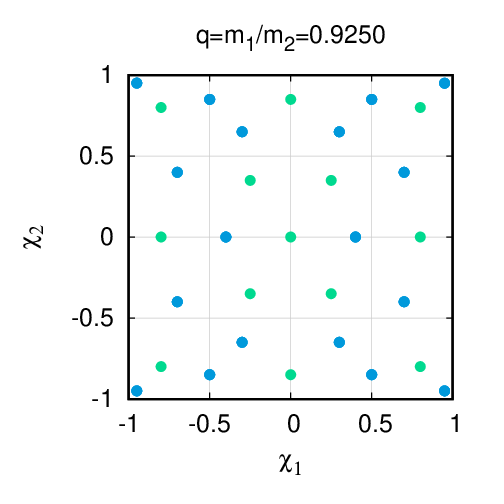}
  \includegraphics[angle=0,width=0.494\columnwidth]{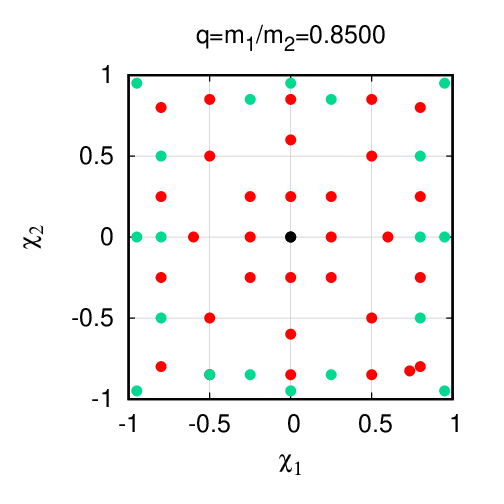}
      \includegraphics[angle=0,width=0.494\columnwidth]{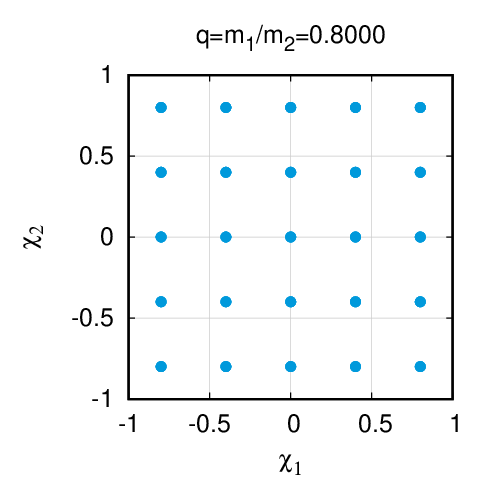}\\
    \includegraphics[angle=0,width=0.494\columnwidth]{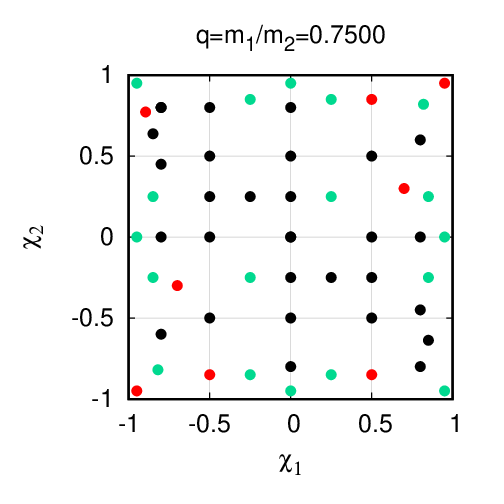}
    \includegraphics[angle=0,width=0.494\columnwidth]{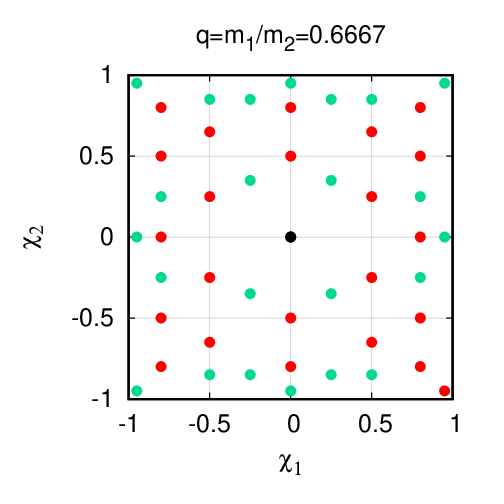}
    \includegraphics[angle=0,width=0.494\columnwidth]{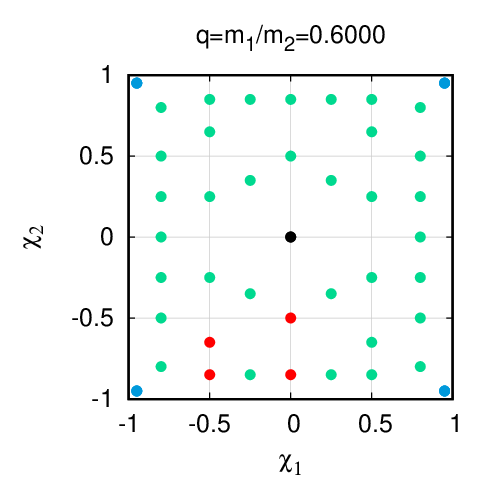}
    \includegraphics[angle=0,width=0.494\columnwidth]{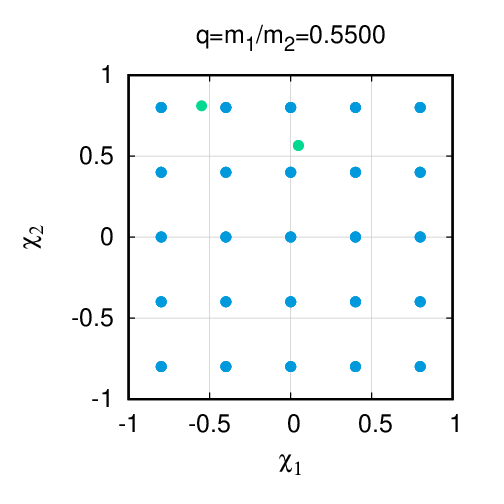}\\
        \includegraphics[angle=0,width=0.494\columnwidth]{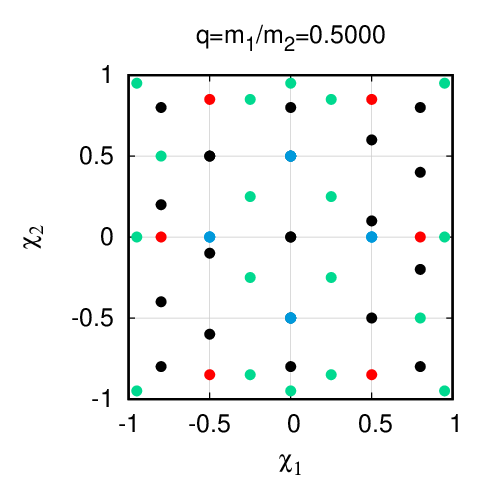}
  \includegraphics[angle=0,width=0.494\columnwidth]{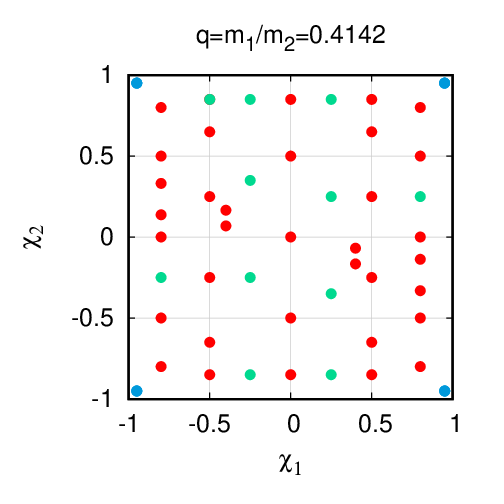}
    \includegraphics[angle=0,width=0.494\columnwidth]{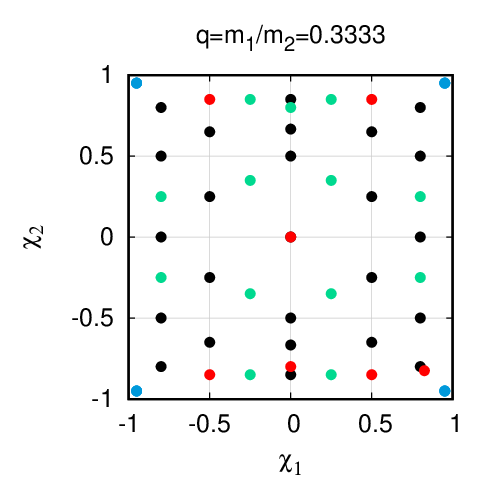}
    \includegraphics[angle=0,width=0.494\columnwidth]{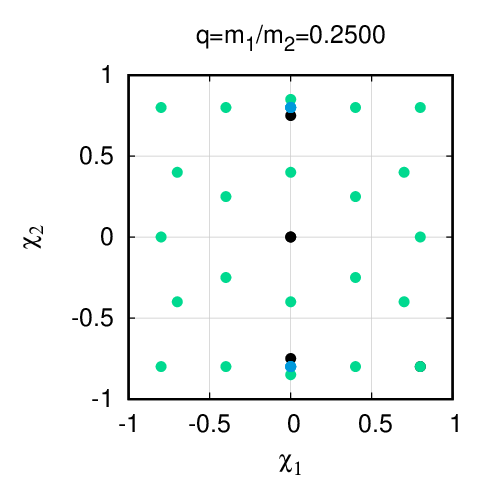}\\
    \includegraphics[angle=0,width=0.494\columnwidth]{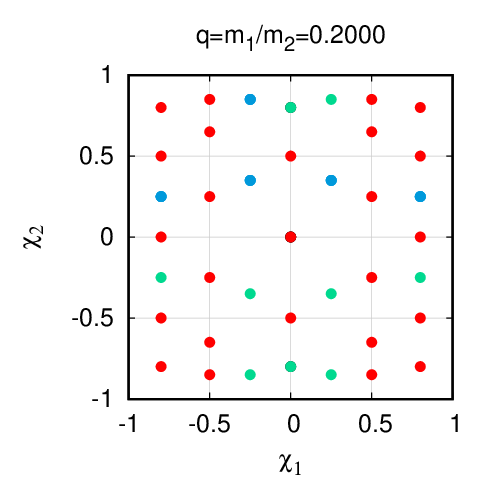}
    \includegraphics[angle=0,width=0.494\columnwidth]{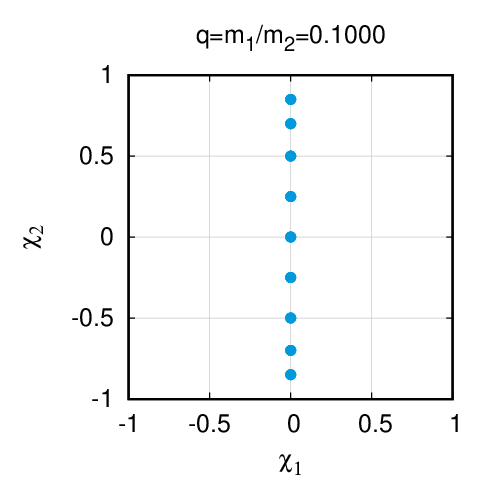}
    \includegraphics[angle=0,width=0.494\columnwidth]{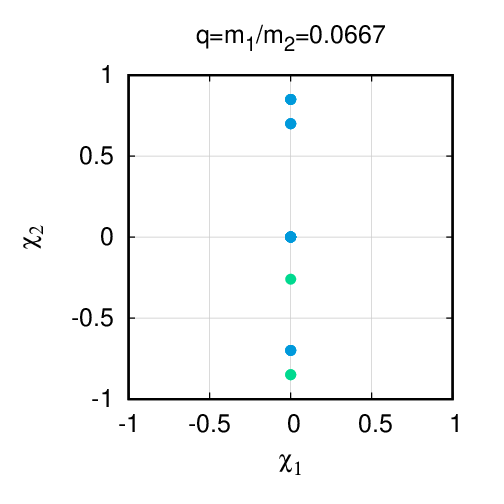}
     \includegraphics[angle=0,width=0.494\columnwidth]{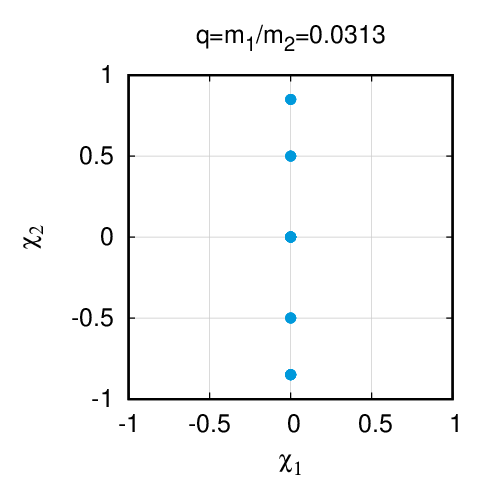}
  \caption{Initial parameters in the $(q,\chi_1,\chi_2)$ space
    for the 611 quasicircular
    nonprecessing binaries. Note that $\chi_i$ denotes 
    the component of the dimensionless spin of BH $i$ along the
    orbital angular momentum.
    Each panel corresponds to a given mass ratio that covers the 
    comparable masses binary range from $q=1$ to $q=1/32$
    (the nonspinning $q=1/64,1/128$) not shown here).
    The dots in black denote the simulations of the catalog first
    release, the dots in red are those of the second release,
    the dots in green are those of this third release,
    and the dots in blue are those of this fourth release.
    \label{fig:panels}}
\end{figure*}
%\end{widetext}

%\begin{widetext}
\begin{figure}[h!]
  \includegraphics[angle=0,width=1.0\columnwidth]{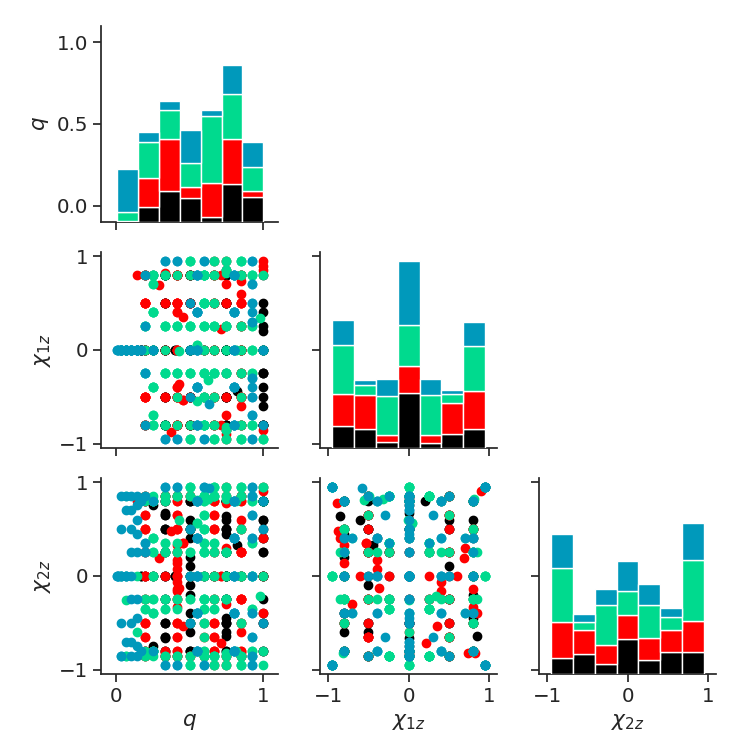}
  \caption{Counting simulations in the $(q,\chi_1,\chi_2)$ planes 
    (faces of the cube) for the 611 nonprecessing binaries.  The 120 release 1
    simulations are black, the 154 release 2 simulations are red, the 203 release 3 are in green, and the 134 release 4 are in blue.
      \label{fig:AlignedMulti}}
\end{figure}
%\end{widetext}

The RIT Catalog can be found at \url{http://ccrg.rit.edu/~RITCatalog}.
Figure~\ref{fig:panels} shows the distribution of the
non-precessing runs in the catalog in
terms of $\chi_{1,2}$ and $q$ (where $\chi_i$ is the component of
the dimensionless spins of BH $i$ along the direction of the orbital angular
momentum).
The information currently in the catalog consists of the metadata
describing the runs and all modes up through the $\ell=4$ modes
of $m r \psi_4$ extrapolated to $\scri^+$ via the perturbative approach
of~\cite{Nakano:2015pta}. 
The associated metadata include the
initial orbital frequencies, ADM masses, initial waveform frequencies from (2,2) mode,  black hole masses, momenta, spins, separations,
and eccentricities, as well the black-hole masses and spins once the
initial burst of radiation has left the region around the binary.
{\it Relaxed} quantities (at $t_{relax}=200m$ after the initial burst
of radiation has mostly dissipated)
are more accurate and physically relevant for modeling purposes. 
We normalize our data such that
the sum of the two initial horizon masses is $1m$.
In addition, we also include peak luminosities, amplitude, and frequency,
and the final remnant black hole masses, spins, and recoil velocities.

The catalog is organized using an interactive table~\cite{datatables_web} that includes an
identification number, resolution, type of run (nonspinning, aligned spins,
precessing), the initial proper length of the coordinate 
line joining the two BH centroids between the two horizons~\cite{Lousto:2013oza},
the coordinate separation of the two centroids, the mass ratio of the two black
holes, the components of the dimensionless spins of the two black
holes, the starting waveform frequency, $m f_{22, {\rm relax}}$, 
time to merger, number of gravitational wave cycles 
calculated from the (2,2) modes from the beginning of the inspiral signal 
to the amplitude peak, remnant mass, remnant spin, recoil
velocity, and peak luminosity. The final column gives 
bibtex keys for the relevant publications where the waveforms were
first presented. The table can be sorted (ascending or descending) by
any of these columns, and there is a direct search feature that runs over all
table elements.
Resolutions are given in terms of the grid spacing of the refinement
level where the waveform is extracted (which is typically two
refinement levels below the coarsest grid, where external boundaries of
the simulations lies) with
the observer location typically about $R_{obs}\sim100m$.
We use the notation 
nXYY, where the grid spacing in the wavezone is given by $h=m/X.YY$, e.g.,
n120 corresponds to $h=m/1.2$, n140 corresponds to $h=m/1.4$, and so on.

For each simulation in the catalog there are three files: one contains
the metadata information in ASCII format, the other two are a tar.gz files
containing ASCII files with up to and including $\ell=4$
modes of $m r\psi_4$ and $H$.
The primary data in our catalog is the Weyl scalar $mr \psi_4$ extrapolated
to $\scri^+$  (using Eq.~(29) of Ref.~\cite{Nakano:2015pta}),
rather than the strain $(r/m)H$, that is also provided as a double time
integration of $mr \psi_4$.

Figure~\ref{fig:AlignedMulti} shows a histogram of the
distribution of the 611 non-precessing
runs in the catalog in terms of $\chi_{1,2}$ and $q$.
Those runs were  motivated by
systematic studies to produce a set of accurate remnant
formulas to represent the final mass, spin and recoil of a merged
binary black hole system and the peak luminosity, amplitude and
frequency, as a function of the parameters of the
precursor binary, as reported in 
\cite{Healy:2014yta,Healy:2016lce,Keitel:2016krm,Healy:2018swt,Varma:2018aht,Varma:2019csw,Healy:2020vre}.
Another important motivation was to provide a grid of simulations for
parameter estimation of gravitational wave signals detected by LIGO
using the methods described in \cite{Abbott:2016apu,Healy:2020jjs}. 

The precessing quasicircular runs in the catalog were motivated to study  
particular spin dynamics of merging BHB, such as
the study of unstable spin flip-flop \cite{Lousto:2016nlp},
beaconing (L-flip) \cite{Lousto:2018dgd},
and targeted followups
of gravitational wave signal from the first and second LIGO observing runs
\cite{Lovelace:2016uwp,Healy:2017abq,Healy:2020jjs}.
We have supplemented here the 300 precessing
simulations of the third catalog release with
additional 146 simulations that improve the coverage of spin orientations
(see Fig.~\ref{fig:precInit}).

\begin{figure}[h!]
\includegraphics[width=0.75\columnwidth]{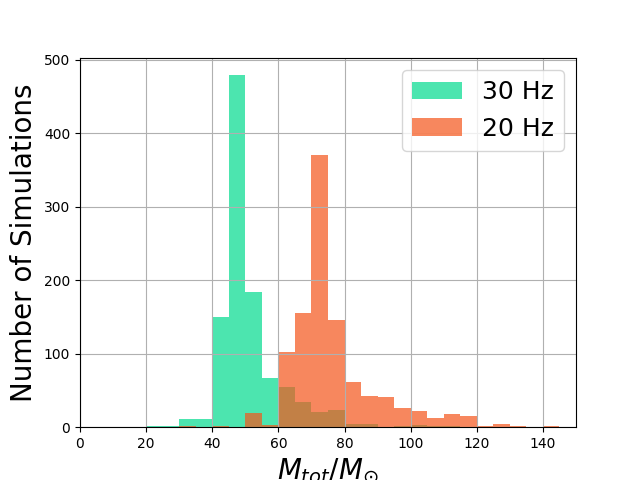}\\
  \includegraphics[width=0.75\columnwidth]{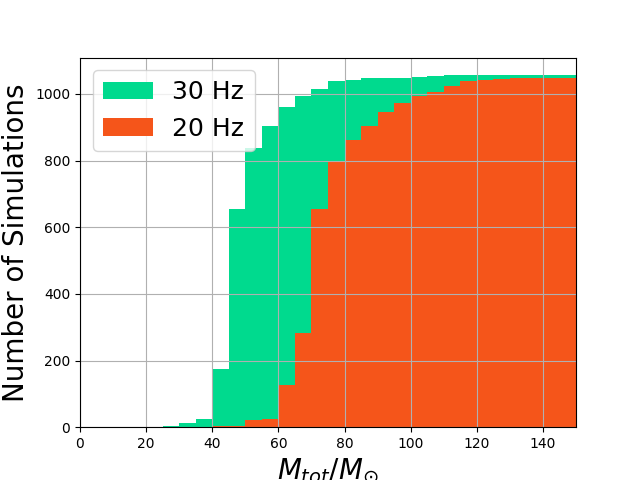}
  \caption{Top: Distributions of the total mass of BHB systems in the
RIT catalog corresponding to a starting gravitational wave frequency of
20 Hz (green) and 30 Hz (red) in bins of $5M_\odot$. 
Bottom: The cumulative version of the above plot also in bins of
$5M_\odot$ for the 1057 quasicircular simulations in this catalog.
\label{fig:mtotal}}
\end{figure}

Figure~\ref{fig:mtotal} shows the distributions of the minimal total mass of the 
1047 quasicircular BHB systems in this fourth catalog release
given a starting gravitational wave frequency
of 20 or 30 Hz in the source frame
(redshift effects improve this coverage in the detector frame
by a factor of $1/(1+z)$ in frequency, where $z$ is the redshift).

%%%%%%

\section{Extension to eccentric merging binaries}\label{sec:eccentricity}

Three body encounters and accretion effects
(see, e.g.,\cite{Schnittman:2015eaa, Nixon:2012zb, Antonini:2012ad, Samsing:2013kua}) 
can lead to highly eccentric binaries, with residual eccentricity surviving down to
merger, and these eccentric binaries may have very interesting
gravitational waves signals that cannot 
be adequately modeled using quasicircular approximations. 
This subject has been the focus of great interest lately
\cite{ShapiroKey:2010cnz,DOrazio:2018jnv,Hoang:2019kye}, 
but its detailed modeling is largely incomplete.
Reliable evolutions between LISA, LIGO bands can also be used to 
exploit multiband observational opportunities
\cite{Sesana:2016ljz,Vitale:2016rfr,Barausse:2016eii}.

In Ref.~\cite{Gayathri:2020coq} we show that GW190521 is most
consistent with a highly eccentric black hole merger. We carried out
619 numerical relativity eccentric simulations to generate an effective
$\sim6\times10^4$ gravitational waveforms with different total masses
to compare to the observed
data. We found that GW190521 is best explained by a
high-eccentricity, precessing model with $e\sim0.7$ at the start
of the simulation. All properties of
GW190521 point to its origin being the repeated gravitational capture
of black holes, making GW190521 the first of LIGO/Virgo's discovery
whose formation channel is identified.
We carried out eccentric binary black hole simulations in this
study, with eccentricities in the full $e \in [0,1]$ range. These
simulations included non-spinning, aligned-spin and spin-precessing
waveforms, and mass ratios $1/7 \leq q=m_2/m_1 \leq 1$. We first
carried out a thorough survey of the eccentricity-mass ratio
parameter space with non-spinning simulations. Then we carried out
aligned/anti-aligned spin and precessing simulations for a broad range
of eccentricity values, most densely covering the parts of the
parameter space where the non-spinning simulations produced the
highest likelihood $\log{\cal L}_{\rm marg}$ in comparison to GW190521
signal. For
precessing waveforms we most densely targeted the $\chi_{\rm p}\sim
0.7$ case expected from black hole merger remnants.  Our simulated
gravitational waveforms were then scaled to correspond to different
black hole masses, providing about 100 scaled mass values for each
simulation in order to best match the total mass of the GW190521
event.

Here we report on 632 runs with a reference quasicircular gravitational wave frequency of 10Hz for a total system mass of $50M_\odot$ (motivated
by the above mentioned target study for GW190521). It includes
115 precessing eccentric binaries and 517 nonprecessing of which
319 are nonspinning covering mass rations from $q=1/128$ to $q=1$;
and 198 have aligned (or counteraligned) spins with the orbital angular momentum. In particular we include here the nonspinning binaries with
mass ratios $q=1/32, 1/64, 1/128$ studied in \cite{Rosato:2021jsq}.

In addition we performed another set of eccentric binaries studies
with a quasicircular gravitational wave reference frequency of 
30Hz for a total system mass of $50M_\odot$ corresponding to
192 nonspinning binaries bearing mass ratios $q=1,0.75,0.50,0.25$.

The numerical simulations techniques described in Sec.~\ref{sec:FN}
also apply to accurately evolve eccentric binaries. 
To compute the numerical initial data, we use the puncture
approach~\cite{Brandt97b} along with the {\sc  TwoPunctures}
~\cite{Ansorg:2004ds} code.  For each eccentric family, we 
first determine the initial separation and tangential quasicircular momentum%~\cite{Healy:2017zqj}
, $p_{t,qc} $, that leads to a frequency of 10\,Hz for a
$50\,$M$_\odot$ system, using the
post-Newtonian techniques described in~\cite{Healy:2017zqj}.  
To increase the eccentricity of the system while keeping the initial data 
at an apocenter, the initial tangential momentum is modified by parameter, $0 < \epsilon < 1$,
such that $p_t = p_{t,qc} ( 1 - \epsilon )$.  The initial eccentricity is then approximately
$e = 2\epsilon-\epsilon^2$. We plan to extend this definition to higher PN
order to improve the identification of high eccentricities.

\begin{figure}[h!]
  \includegraphics[angle=0,width=1.0\columnwidth]{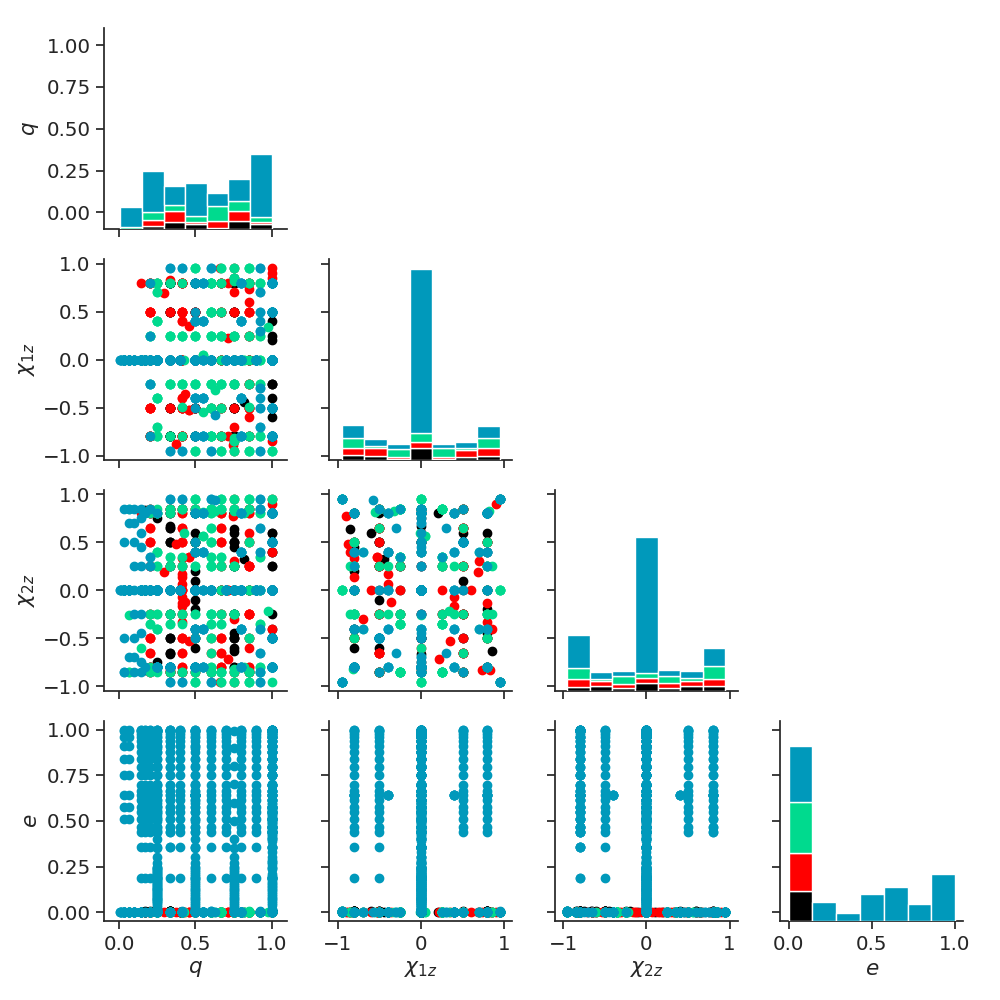}
  \caption{Counting simulations in the $(q,\chi_1,\chi_2,e)$ planes 
    (faces of the 4-cube) for the 1320 nonprecessing binaries.  The 120 release 1
    simulations are black, the 154 release 2 simulations are red, the 203 release 3 are in green, and the 843 release 4 in blue.
      \label{fig:AlignedMulti4}}
\end{figure}

Figure~\ref{fig:AlignedMulti4} shows the distribution of the 
611 quasicircular and 709 eccentric non-precessing
runs in the catalog, per release, in terms of the spins $\chi_{1,2}$,
mass ratios $q$ and eccentricity $e$.

%\begin{widetext}
%\end{widetext}

%%%%%%

\section{merging binaries correlations}\label{sec:correlations}

Detailed and higher order formulas relating the binary parameters to
the post-merger properties of the final remnant black hole and merger
waveform have been studied in 
Refs. \cite{Healy:2014yta,Healy:2016lce,Healy:2018swt} using the
aligned spins simulations of previous releases of the RIT catalog.

\begin{figure*}
\hfill\includegraphics[angle=0,width=0.66\columnwidth]{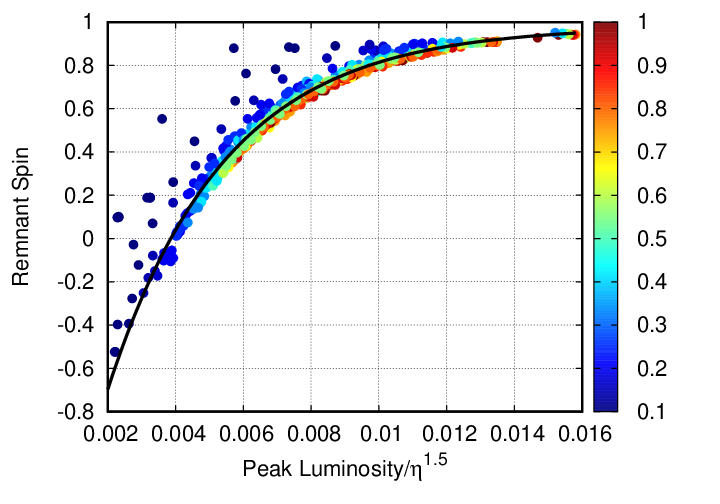}
\includegraphics[angle=0,width=0.66\columnwidth]{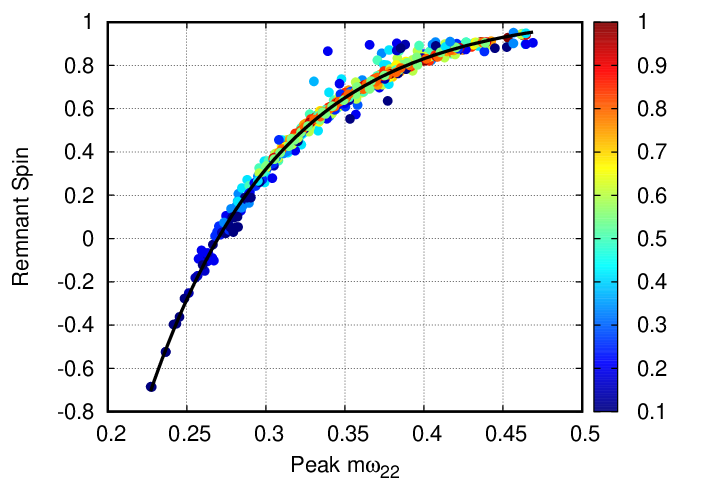}
  \includegraphics[angle=0,width=0.66\columnwidth]{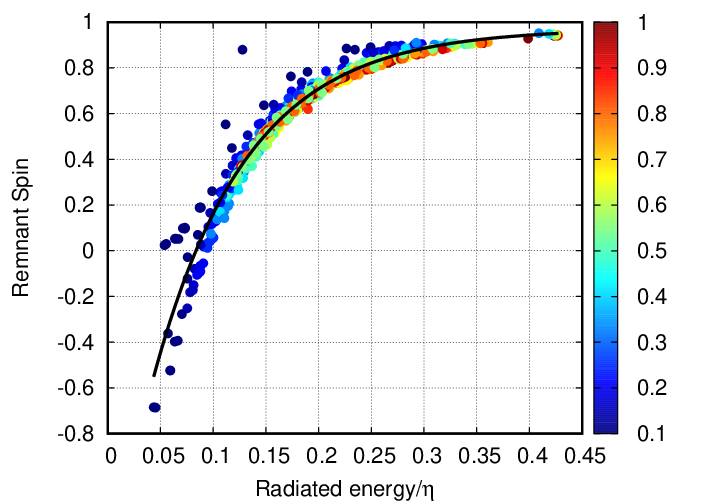}\\
  \hfill\includegraphics[angle=0,width=0.66\columnwidth]{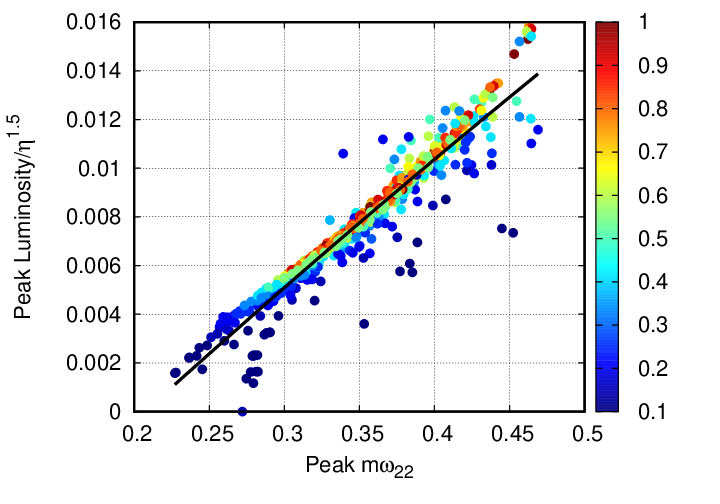}
    \includegraphics[angle=0,width=0.66\columnwidth]{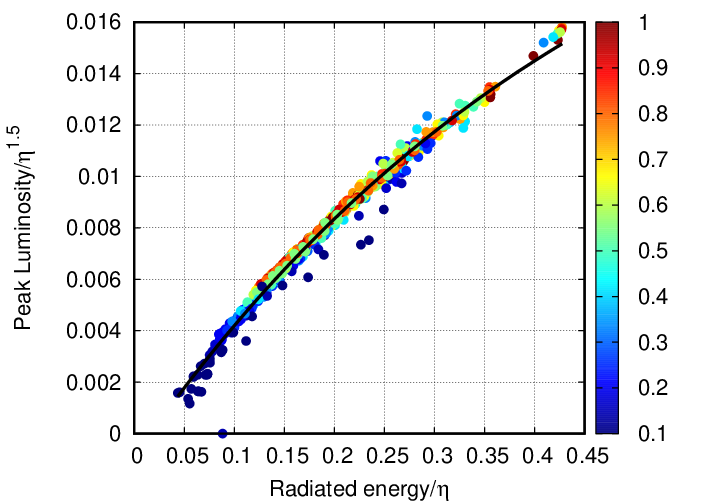}\\
    \quad\ \includegraphics[angle=0,width=0.66\columnwidth]{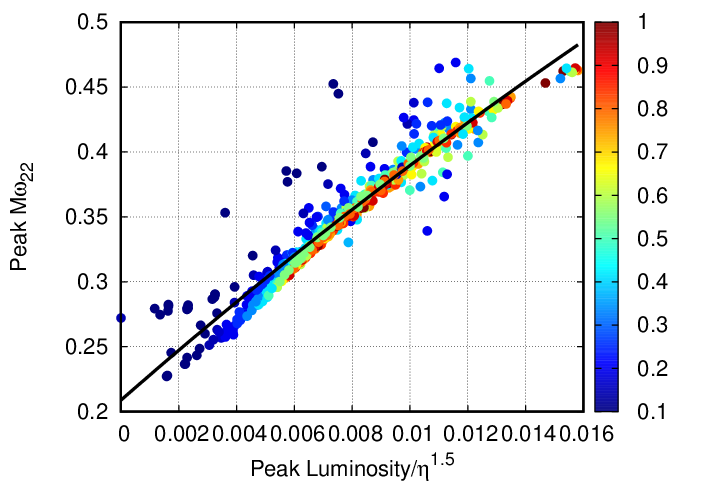}
    \hfill\includegraphics[angle=0,width=0.66\columnwidth]{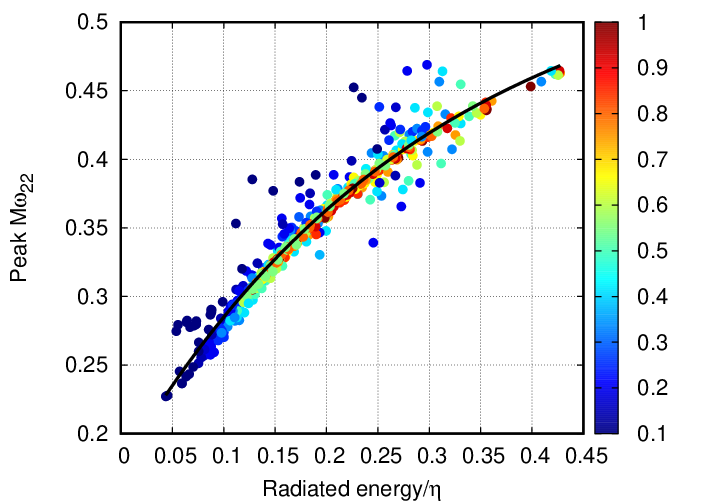}
  \caption{Correlations between the radiated energy, peak luminosity, peak
    frequency at merger and spin remnant with the corresponding leading scaling
    by the symmetric mass ratio. Color column bar corresponds to different mass ratios of these quasicircular simulations.
          \label{fig:correlations}}
\end{figure*}

Foreseeing astrophysical applications of future massive catalogs
of binary black holes,
we display in Fig. \ref{fig:correlations}
simple scaling phenomenological
correlations between remnant parameters and merger waveforms for quasicircular binaries.

%\begin{widetext}
%\end{widetext}

Fitting formulas of the form $f(x) = a + b\cdot\exp(c\cdot x)$
for these simple correlations and estimated errors of
their fitted coefficient are given by
\bea\label{eq:correlations1}
%\begin{align}
&&\alpha_f  = 1.049\pm0.007 - (27.07\pm0.97) \mathrm{e}^{(-12.05\pm0.15) m\omega_{22}}, \nonumber\\
&&\alpha_f  = 0.980\pm0.001 - ( 2.99\pm0.09) \mathrm{e}^{(-288.85\pm7.96) \mathcal{L}/\eta^{1.5}}, \nonumber\\
&&\alpha_f  = 0.971\pm0.009 - ( 2.48\pm0.04) \mathrm{e}^{(-11.21\pm0.25) E_{rad}/\eta},\nonumber\\
%\end{align}
\eea
for the final spin $\alpha_f$ as a function of the peak frequency
$m\omega_{22}$ of the (2,2) mode, the peak Luminosity $\mathcal{L}$,
and the total radiated energy during merger $E_{rad}$
(normalized by the symmetric mass ratio $\eta=q/(1+q)$).
The typical error in the fitted coefficients amount to less than $3\%$.
Note that the application of the fits should be for intermediate
mass ratios. In particular the spin versus small $E_{rad}$ and $\mathcal{L}$ will
be dominated in the small mass ratio regime by the (original) spin of the large
hole and hence the observed branching off of curves at those small values.

Similar correlations among other quantities are,
\bea\label{eq:correlations2}
%\begin{align}
&&\mathcal{L}/\eta^{1.5}  = 0.027\pm0.001 - (0.027\pm0.001)\mathrm{e}^{(-2.05\pm0.10) E_{rad}/\eta},\nonumber\\
&&\mathcal{L}/\eta^{1.5}  = 0.138\pm0.136 - (0.151\pm0.135)\mathrm{e}^{(-0.40\pm0.42) m\omega_{22}},\nonumber\\
&&m\omega_{22}  = 0.562\pm0.013 - (0.38\pm0.01) \mathrm{e}^{(-3.32\pm0.21) E_{rad}/\eta}.\nonumber\\
&&m\omega_{22}  = 1.487\pm0.048 - (1.28\pm0.48) \mathrm{e}^{(-15.24\pm6.50) \mathcal{L}/\eta^{1.5}}\nonumber\\
%\end{align}
\eea

%The fittings take the leading behavior and scaling of the correlations
%without attempting higher order corrections to be used in astrophysical
%estimates and control of more sophisticated implementations in large
%catalogs of binary black hole gravitational wave signals and its modeling,
%as well as tests of gravity.

%%%%%%%

In the case of the eccentric binaries we will seek for similar
correlations that also involve only the non precessing simulations
for which we have a set of 709 simulations, all in this fourth catalog release.

%for precessing as well as non precessing use the whole set
%of 824 simulations in this fourth catalog release.

\begin{figure*}
\hfill
\includegraphics[angle=0,width=0.66\columnwidth]{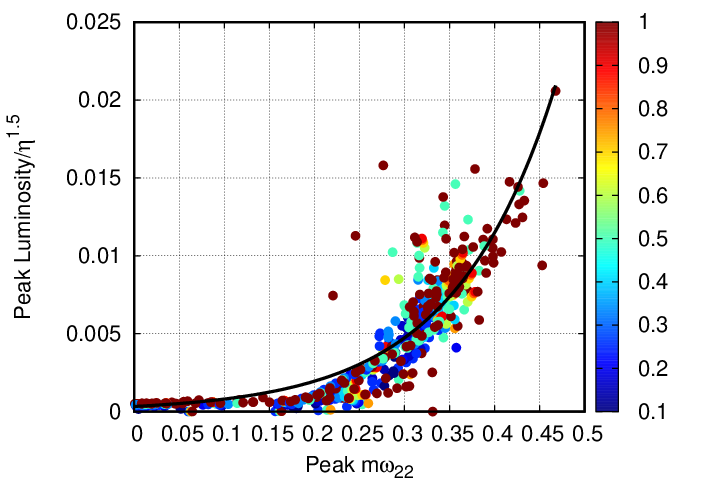}
\includegraphics[angle=0,width=0.66\columnwidth]{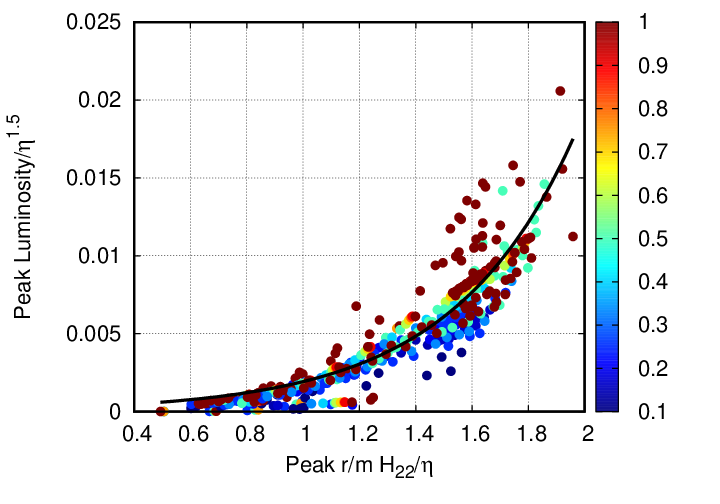}
\includegraphics[angle=0,width=0.66\columnwidth]{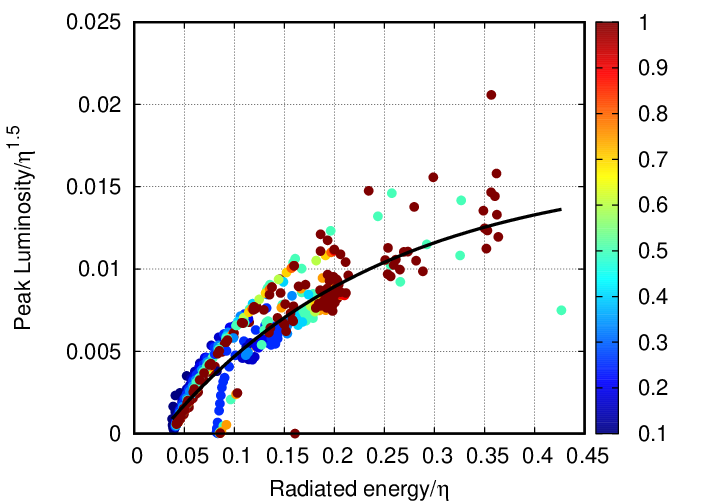}\\
\hfill
\includegraphics[angle=0,width=0.66\columnwidth]{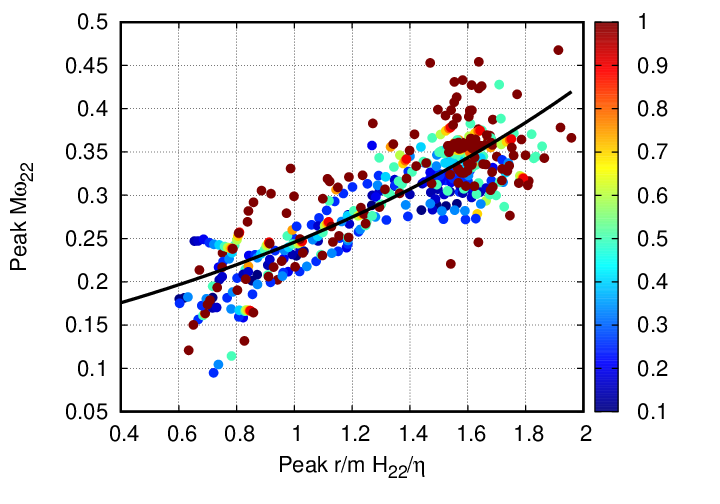}
\includegraphics[angle=0,width=0.66\columnwidth]{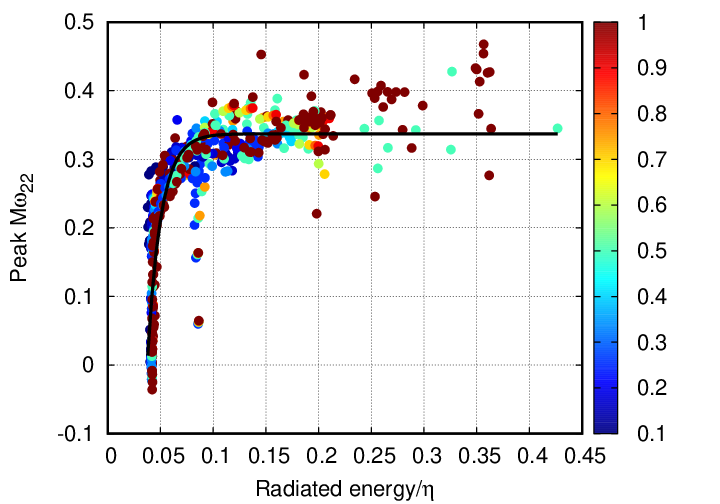}\\
\hfill
\includegraphics[angle=0,width=0.66\columnwidth]{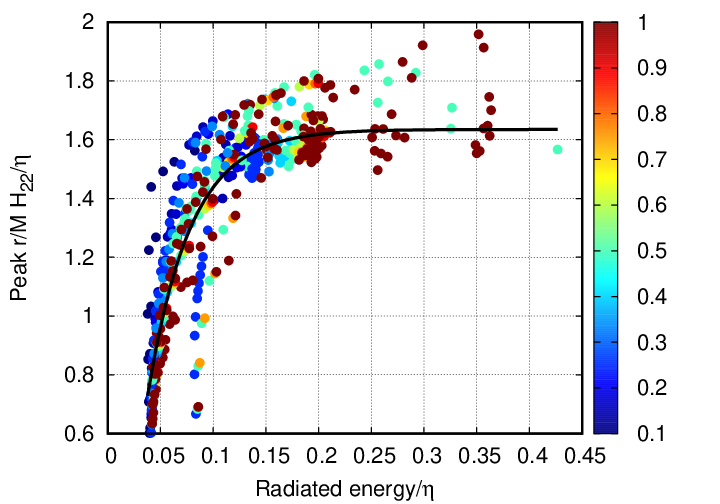}\\
  \caption{Phenomenological correlations for eccentric binaries between the radiated energy, peak luminosity, peak frequency at merger
    with the corresponding scaling by the symmetric mass ratio $\eta$
    to obtain the leading dependences. The color column bar on the right corresponds to different mass ratios.
    \label{fig:ecorrelations}}
\end{figure*}

Fitting formulas for the simple correlations and estimated errors in
their coefficient are 
\bea\label{eq:ecorrelations1}
%\begin{align}
&&\mathcal{L}/\eta^{1.5}  = (0.00033\pm0.000026)\mathrm{e}^{(8.87\pm0.22) m\omega_{22}}.\nonumber\\
&&\mathcal{L}/\eta^{1.5}  = (0.00019\pm0.000016)\mathrm{e}^{(2.30\pm0.05) (r/m)H_{22}/\eta}.\nonumber\\
&&\mathcal{L}/\eta^{1.5}  = (0.0162\pm0.001)-(0.018\pm0.0008)\mathrm{e}^{(-4.58\pm0.47) E_{rad}/\eta},\nonumber\\
%\end{align}
\eea

for the final peak luminosity $\mathcal{L}$ as a function of the peak frequency
$m\omega_{22}$ and waveform amplitude $H_{22}$ (extrapolated to $\scri^+$) of the (2,2) mode,
and the total radiated energy during merger $E_{rad}$
(normalized by the symmetric mass ratio $\eta=q/(1+q)$).

Other correlations among those quantities can be found,
\bea\label{eq:ecorrelations2}
%\begin{align}
&&m\omega_{22}=(0.141\pm0.003)\mathrm{e}^{(0.558\pm0.016)\frac{r}{m}H_{22}/\eta}, \nonumber\\
&&m\omega_{22}=(0.357\pm0.003)-(0.385\pm0.0027)\mathrm{e}^{(-22.69\pm1.70)E_{rad}/\eta},\nonumber\\
&&(r/m)H_{22}/\eta=(1.635\pm0.011)-(2.29\pm0.14)\times\nonumber\\
&&\quad\quad\quad\quad\quad\quad\quad\quad\quad\quad\quad\quad\quad\quad\times\mathrm{e}^{(-24.52\pm1.41)E_{rad}/\eta}.
%\end{align}
\eea

These correlations for highly eccentric orbits carry larger errors than
the corresponding correlations for quasicircular/inspiral orbits, but
still could be used in astrophysical estimates. Their range of validity
should be for intermediate mass ratios and in their positive intervals,
where these simple formulas have been fitted and we have not tested
them in the extrapolation regime.

%%%%%%%%%%%%%%%%%%%%%%%%%%%%%%%%%%%%%%%%

\section{Conclusions and Discussion}\label{sec:Discussion}

The 2005 numerical relativity
breakthroughs~\cite{Pretorius:2005gq,Campanelli:2005dd,Baker:2005vv}
 were instrumental in identifying the first
detection of gravitational waves \cite{TheLIGOScientific:2016wfe} with
the merger of two black holes.
Those different approaches to solve the binary black hole problem
produced an excellent agreement for the modeling of the sources of
GW150914~\cite{Lovelace:2016uwp} and GW170104~\cite{Healy:2017abq},
including comparison of higher (up to $\ell=5$) modes.
The recent success of applying eccentric orbits simulations to
describe the source of GW190521~\cite{Gayathri:2020coq}
highlights the importance of the direct use of numerical relativity
waveforms to model sources of gravitational waves events.

In particular, we used the third RIT Catalog release \cite{Healy:2020vre}
to reanalyze the ten binary black hole events reported by the
LIGO-Virgo collaboration for the O1/O2 observational runs
\cite{LIGOScientific:2018mvr}, confirming,
and sometimes improving their parameter estimations~\cite{Healy:2020jjs}.
We have also succeeded in confirming the parameters of three additional
black hole binary merger events, displaying again the importance of
the use of numerical relativity waveform catalogs as a consistent
method for parameter estimation.

The next areas of development for the numerical relativity waveform
catalogs include the coverage of precessing binaries. Those require
expansion of the parameter space to eight dimensions (including now
eccentricity), and can be carried out in a hierarchical
approach by first neglecting the effects of the spin of the secondary black
hole, which is a good assumption for small mass ratios.
Other lines of extension of the catalogs include very highly spinning
black holes with spin magnitudes in the range $\chi=0.95-0.99$.
And smaller mass ratio binaries to complete the family of simulations displayed
in Fig. \ref{fig:panels}, i.e. $q=1/10,1/15,1/32$ and include spins of
the large hole in the $q=1/64,1/128$ simulations. These small mass
ratio simulations, in turn, can inform the modeling through phenomenological
approaches, like the use of effective one body (EOB) models.%~\cite{Nagar:2022}.
While low (below $20M_\odot$) total binary masses, require longer
full numerical simulations or hybridization of the current numerical relativity waveforms
with post-Newtonian waveforms \cite{Sadiq:2020hti}.

Simulations of black-hole binaries also produce
information about the final remnant of the merger.
Several empirical formulas relating the initial parameters
$(q,\vec\chi_1,\vec\chi_2)$ (individual masses and spins) of the
binary to those of the final remnant $(m_f,\vec\alpha_f,\vec{V}_f)$ have
been proposed for the final mass, spin, and
recoil velocity
\cite{Barausse:2012qz,Rezzolla:2007rz,Hofmann:2016yih,Jimenez-Forteza:2016oae,Lousto:2009mf,Lousto:2013wta,Hemberger:2013hsa,Healy:2014yta,Zlochower:2015wga,Gerosa:2018qay,Varma:2019csw},
the computation of the peak frequency of the (2,2) mode
$\Omega_{22}^{peak}$, peak waveform
amplitude $H_{22}^{peak}$
~\cite{Healy:2017mvh,Healy:2018swt} and peak luminosity
\cite{TheLIGOScientific:2016wfe,TheLIGOScientific:2016pea,Healy:2016lce,Keitel:2016krm}.
The tables in the Appendix \ref{app:ID}
can be used to further model those remnant formulas and to include
eccentricity as an additional parameter
and to test remnant and merger waveform
parameters in terms of those of the inspiraling binary as a consistency test
for the theory of gravity.

We finally summarize here the number of simulations per release in
Table~\ref{tab:Catalogs} with a distinction of quasicircular
(QC) and eccentric orbits and by discriminating binary black holes
on being both nonspinning or spinning but non precessing, and
full precessing.

\begin{table}
\caption{Summary of simulations in each release of the RIT catalog}
\label{tab:Catalogs}
\begin{ruledtabular}
\begin{tabular}{|l|c|c|c|c|c|}
  Release / [cite] & \#1~\cite{Healy:2017psd} & \#2~\cite{Healy:2019jyf} & \#3~\cite{Healy:2020vre} & \#4~%\cite{}
  & Total\\
\hline
QC Nonspinning & 21 & 4 & 4 & 14 & 43 \\
QC Aligned  & 99 & 150 & 199 & 120 & 568 \\
%QC Aligned  & 120 & 154 & 203 & 134 & 611 \\
QC Precessing  & 6 & 40 & 254 & 146 & 446 \\
Eccentric Nonspinning & - & - & - & 511 & 511 \\
Eccentric Aligned  & - & - & - & 198 & 198 \\
Eccentric Precessing  & - & - & - & 115 & 115 \\
\hline
Subtotal   & 126 & 194 & 457 & 1104 & 1881 \\
\end{tabular}
\end{ruledtabular}
\end{table}

\begin{figure*} [htp]
\includegraphics[angle=0,width=2.00\columnwidth]{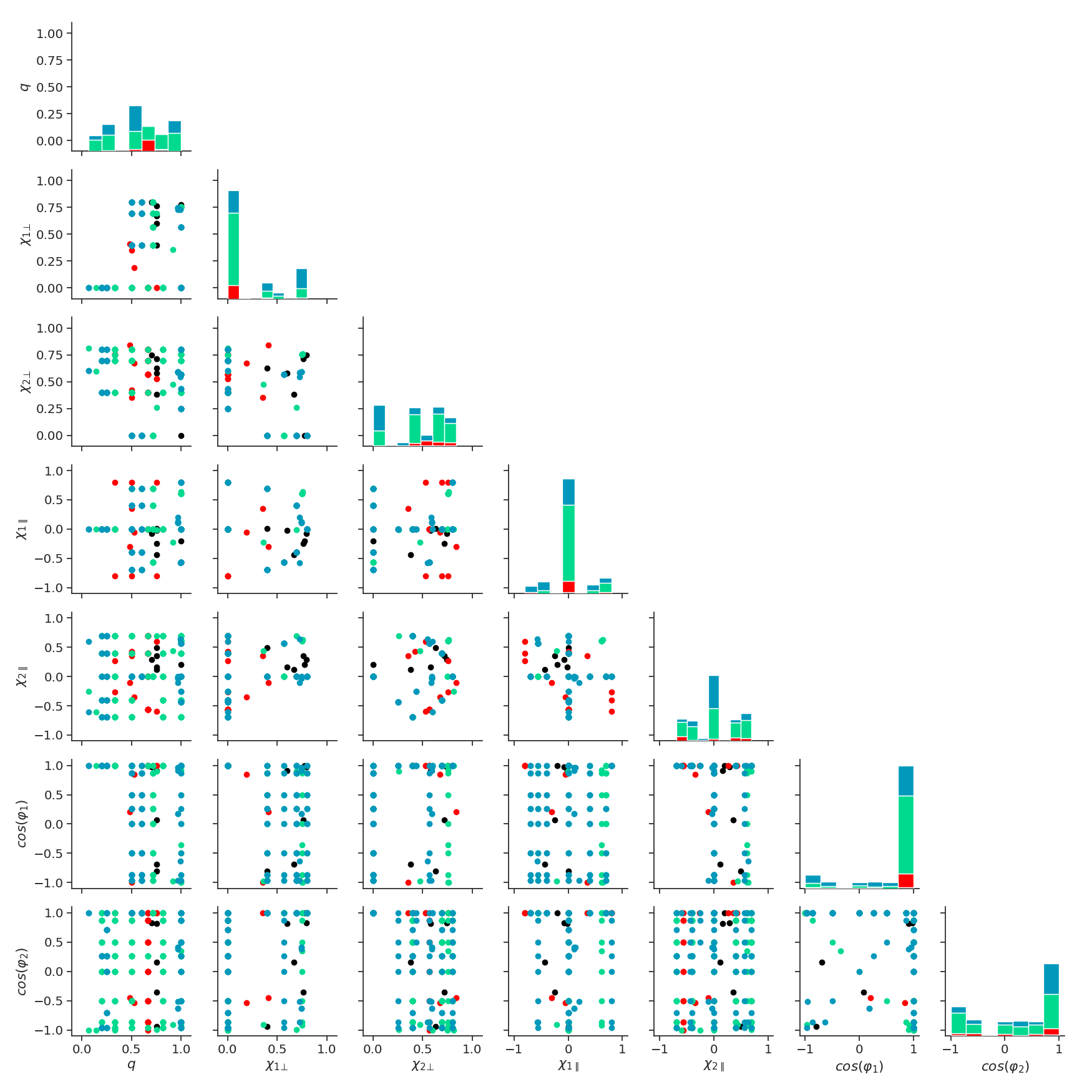}
\caption{Panels showing different combinations of the 7 binary parameters of the precessing parameter space $(q, |\chi_1|, \theta_1, \varphi_1, |\chi_2|, \theta_2, \varphi_2)$ for the 446 quasicircular simulations (black first release, red second release, green third release, blue fourth release) in this catalog.
\label{fig:precInit}}
\end{figure*}

A visual representation of the 446 binaries' in the 
7 QC parameter precessing space is displayed in Fig.~\ref{fig:precInit}.

\begin{figure*}[htp]
\includegraphics[angle=0,width=2.00\columnwidth]{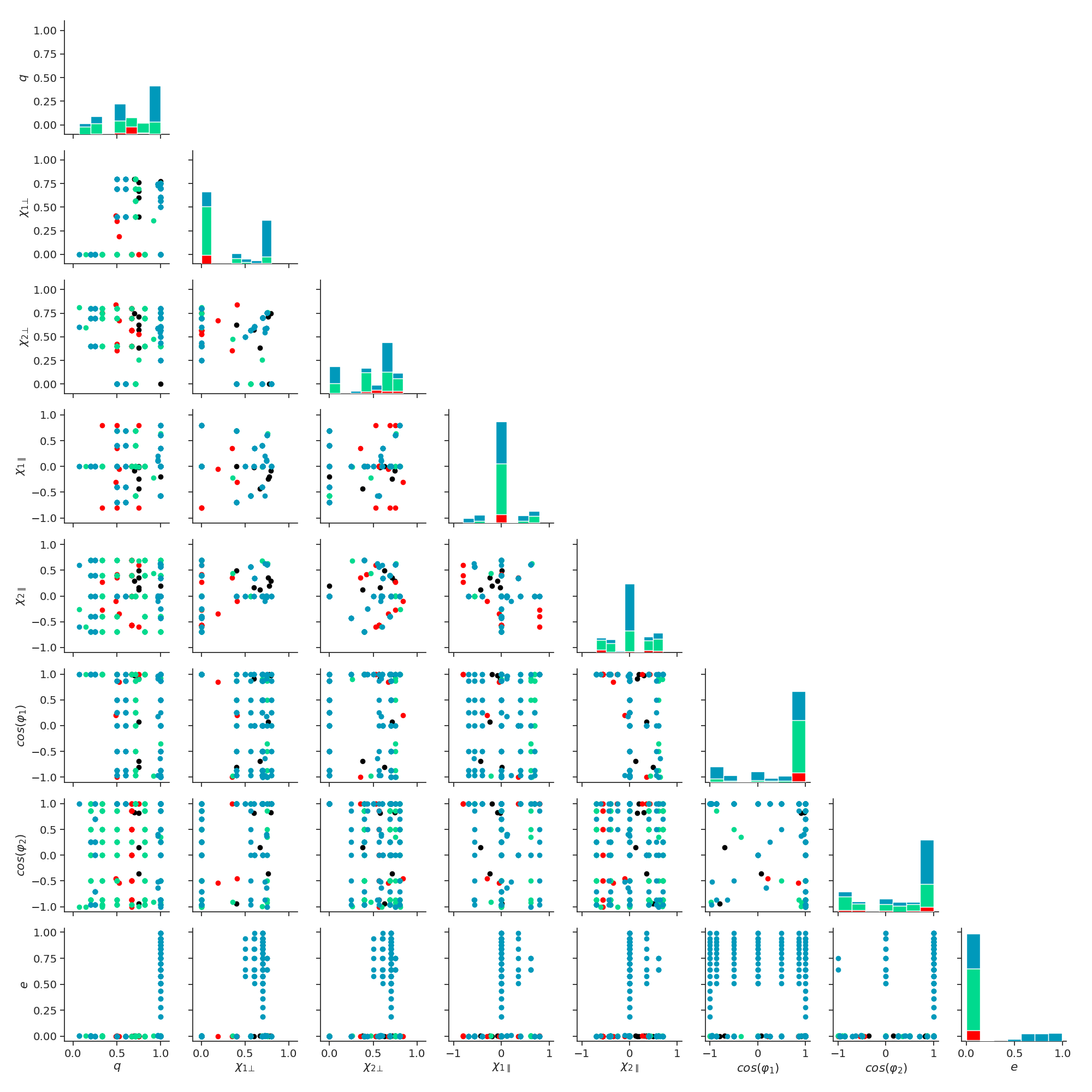}
\caption{Panels showing different combinations of the 8 binary parameters of the precessing parameter space $(q, |\chi_1|, \theta_1, \varphi_1, |\chi_2|, \theta_2, \varphi_2,e)$ for the 561 eccentric simulations (black first release, red second release, green third release, blue fourth release) in this catalog.
\label{fig:precInit8}}
\end{figure*}

While a visualization of the 561 highly eccentric binaries' in the
8 parameter precessing space is displayed in Fig.~\ref{fig:precInit8}.

%%%%%%%%%%%%%%%%%%%%%%%%%%%%%%%%%%%%%%%%%%%%%

\begin{acknowledgments}
 The authors gratefully acknowledge the National Science Foundation
(NSF) for financial support from Grant No.\ PHY-1912632.
Computational resources were also provided by the NewHorizons, BlueSky
Clusters, Green Prairies, and White Lagoon at the Rochester Institute
of Technology, which were supported by NSF grants No.\ PHY-0722703,
No.\ DMS-0820923, No.\ AST-1028087, No.\ PHY-1229173,
No.\ PHY-1726215, and No.\ PHY-2018420.  This work used the Extreme
Science and Engineering Discovery Environment (XSEDE) [allocation
  TG-PHY060027N], which is supported by NSF grant
No.\ ACI-1548562. Frontera is an NSF-funded petascale computing system
at the Texas Advanced Computing Center (TACC).
\end{acknowledgments}

%%%%%%%%%%%%%%%%%%%%%%%%%%%%%%%%%%%%%%%%

\bibliographystyle{apsrev4-1}
\bibliography{../../Bibtex/references}

%%%%%%%%%%%%%%%%%%%%%%%%%%%%%%%%%%%%%%%%%%%%
% Suppressing appendix
%\end{document}
%%%%%%%%%%%%%%%%%%%%%%%%%%%%%%%%%%%%%%%%%%%%
%\newpage

\appendix*
\section{Tables of initial data and results of the new simulations}\label{app:ID}

In this appendix we provide tables with the relevant BBH
configuration details.  In Table \ref{tab:qcID}, we provide the
initial data parameters for the new 280 quasicircular configurations
used to start the full numerical evolutions. Those include 14 nonspinning,
120 spin aligned (nonprecessing) and 146 misaligned spins (precessing) binaries.
In Table \ref{tab:eccID}, we provide the
initial data parameters for the new 192 nonspinning eccentric
configurations used to start the full numerical evolutions
for the aphastron by use of the prescription of
decreasing the quasicircular
tangential orbital momentum by a factor $(1-\epsilon)$. These new simulations
use as a reference starting gravitational wave frequency 30Hz for a
50 $M_\odot$ binary and is an additional set to those remaining 626
with reference frequency 10Hz for a 50 $M_\odot$ binary described in
the supplementary material of Ref.~\cite{Gayathri:2020coq}. %\NOTE{We actually speak about only 611 runs in this paper...non-spinning (313), aligned-spin (37), anti-aligned-spin (123), head-on (35) and spin-precessing (111) waveforms=619?}

In Tables \ref{tab:IDr} and \ref{tab:IDr_prec},
we provide the binary mass and spin parameters after 
they settle into a more physical value after radiating and absorbing
the spurious gravitation wave content from the initial mathematical 
choice of conformal flatness.  These relaxed values are calculated at  
a fiducial $t=200m$.  Table \ref{tab:IDr} includes the new 134 quasicircular
simulations (14 nonspinning) and the new 192 nonspinning eccentric
started at 30Hz gravitational wave frequency for a 50 $M_\odot$ binary.
Table \ref{tab:IDr_prec} adds the new 146 precessing quasicircular
black hole binaries evolutions.

Finally, In Table \ref{tab:spinerad}, we give
the values of the gravitational energy radiated during the simulation
and the final black hole spin as measured through the (accurate)
isolated horizon formalism \cite{Dreyer02a} for all the new simulations
reported in this paper.

%%%%%%%%%%%%%%%%%%%%%%%%%%%%%%%%%%%%%%%%%%%%
%\clearpage

%\clearpage
\begin{longtable*}{lccccccccccccc}
\caption{Initial data parameters for the 280 quasi-circular
configurations with a smaller mass black hole (labeled 1),
and a larger mass spinning black hole (labeled 2). The punctures are located
at $\vec r_1 = (x_1,0,0)$ and $\vec r_2 = (x_2,0,0)$, with individual linear momenta
$P=\pm (P_r, P_t,0)$, spin magnitudes $|S_i|$, puncture mass parameters
$m^p/m$, horizon (Christodoulou) masses $m^H/m$, total ADM mass
$M_{\rm ADM}$, and dimensionless spins $|a/m_H| = |S/m_H^2|$.
The spin directions for the precessing simulations are given in
the catalog and in Tab.~\ref{tab:IDr_prec}.}
\label{tab:qcID}\\
\hline
\hline
Run   & $x_1/m$ & $x_2/m$  & $P_r/m$    & $P_t/m$ & $m^p_1/m$ & $m^p_2/m$ & $|S_1/m^2|$ & $|S_2/m^2|$ & $m^H_1/m$ & $m^H_2/m$ & $M_{\rm ADM}/m$ & $|a_1/m_1^H|$ & $|a_2/m_2^H|$\\
\hline
\endfirsthead

\multicolumn{14}{c}
{{\tablename\ \thetable{} -- continued from previous
page}}\\
\hline
Run   & $x_1/m$ & $x_2/m$  & $P_r/m$    & $P_t/m$ & $m^p_1/m$ & $m^p_2/m$ & $|S_1/m^2|$ & $|S_2/m^2|$ & $m^H_1/m$ & $m^H_2/m$ & $M_{\rm ADM}/m$ & $|a_1/m_1^H|$ & $|a_2/m_2^H|$\\
\hline
\endhead
\hline \multicolumn{14}{r}{{Continued on next page}} \\
\hline
\endfoot
\hline\hline
\endlastfoot

RIT:BBH:0787 & -7.21 & 4.32 & -5.10e-04 & 0.08076 & 0.2259 & 0.6139 & 0.1125 & 0 & 0.375 & 0.625 & 0.9915 & 0.8 & 0 \\
RIT:BBH:0791 & -10.31 & 0.6875 & -5.05e-05 & 0.02214 & 0.05918 & 0.4931 & 0 & 0.7471 & 0.0625 & 0.9375 & 0.9979 & 0 & 0.85 \\
RIT:BBH:0803 & -10.83 & 2.17 & -1.20e-04 & 0.04416 & 0.1562 & 0.7851 & 0.006944 & 0.2431 & 0.1667 & 0.8333 & 0.9954 & 0.25 & 0.35 \\
RIT:BBH:0804 & -10.83 & 2.17 & -1.20e-04 & 0.04419 & 0.09963 & 0.8065 & 0.02222 & 0.1736 & 0.1667 & 0.8333 & 0.9954 & 0.8 & 0.25 \\
RIT:BBH:0806 & -7.98 & 3.99 & -4.56e-04 & 0.07676 & 0.2005 & 0.6565 & 0.08889 & 0 & 0.3333 & 0.6667 & 0.9924 & 0.8 & 0 \\
RIT:BBH:0807 & -9.98 & 2.03 & -2.04e-04 & 0.04919 & 0.1596 & 0.5147 & 0 & 0.5556 & 0.1667 & 0.8333 & 0.9953 & 0 & 0.8 \\
RIT:BBH:0808 & -9.99 & 2.03 & -2.05e-04 & 0.04923 & 0.1596 & 0.5147 & 0 & 0.5556 & 0.1667 & 0.8333 & 0.9954 & 0 & 0.8 \\
RIT:BBH:0809 & -8.96 & 1.79 & -2.37e-04 & 0.05068 & 0.09877 & 0.8054 & 0.02222 & 0.1736 & 0.1667 & 0.8333 & 0.9947 & 0.8 & 0.25 \\
\end{longtable*}

\begin{longtable*}{lccccccccccc}
\caption{Initial data parameters for the new 192 nonspinning eccentric
configurations  with a smaller mass black hole (labeled 1),
and a larger mass spinning black hole (labeled 2). The punctures are located
at $\vec r_1 = (x_1,0,0)$ and $\vec r_2 = (x_2,0,0)$, with individual linear momenta
$P=\pm (P_r, P_t,0)$, spin magnitudes $|S_i|$, puncture mass parameters
$m^p/m$, horizon (Christodoulou) masses $m^H/m$, total ADM mass
$M_{\rm ADM}$ and initial Newtonian eccentricity. Also given is the 
number of orbits to merger, $N_{orb}$. }
\label{tab:eccID}\\
\hline
\hline
Run   & $x_1/m$ & $x_2/m$  & $P_r/m$    & $P_t/m$ & $m^p_1/m$ & $m^p_2/m$ & $m^H_1/m$ & $m^H_2/m$ & $M_{\rm ADM}/m$ & $e$ & $N_{orb}$\\
\hline
\endfirsthead

\multicolumn{12}{c}
{{\tablename\ \thetable{} -- continued from previous
page}}\\
\hline
Run   & $x_1/m$ & $x_2/m$  & $P_r/m$    & $P_t/m$ & $m^p_1/m$ & $m^p_2/m$ & $m^H_1/m$ & $m^H_2/m$ & $M_{\rm ADM}/m$ & $e$ & $N_{orb}$\\
\hline
\endhead
\hline \multicolumn{12}{r}{{Continued on next page}} \\
\hline
\endfoot
\hline\hline
\endlastfoot

RIT:eBBH:1090 & -5.68 & 5.68 & 0 & 0.08822 & 0.4876 & 0.4876 & 0.5 & 0.5 & 0.9907 & 0 & 8.7 \\
RIT:eBBH:1091 & -5.68 & 5.68 & 0 & 0.08813 & 0.4876 & 0.4876 & 0.5 & 0.5 & 0.9907 & 0.002 & 8.6 \\
RIT:eBBH:1092 & -5.68 & 5.68 & 0 & 0.08804 & 0.4876 & 0.4876 & 0.5 & 0.5 & 0.9907 & 0.004 & 8.5 \\
RIT:eBBH:1093 & -5.68 & 5.68 & 0 & 0.08795 & 0.4876 & 0.4876 & 0.5 & 0.5 & 0.9907 & 0.006 & 8.5 \\
RIT:eBBH:1094 & -5.68 & 5.68 & 0 & 0.08786 & 0.4876 & 0.4876 & 0.5 & 0.5 & 0.9906 & 0.008 & 8.4 \\
RIT:eBBH:1095 & -5.68 & 5.68 & 0 & 0.08778 & 0.4876 & 0.4876 & 0.5 & 0.5 & 0.9906 & 0.01 & 8.4 \\
RIT:eBBH:1096 & -5.68 & 5.68 & 0 & 0.08755 & 0.4876 & 0.4876 & 0.5 & 0.5 & 0.9906 & 0.015 & 8.2 \\
RIT:eBBH:1097 & -5.68 & 5.68 & 0 & 0.08733 & 0.4877 & 0.4877 & 0.5 & 0.5 & 0.9905 & 0.02 & 8.0 \\
\end{longtable*}

\clearpage
\begin{longtable}{cccccc}
\caption{The masses and spins of the (134 + 192) nonprecessing BHBs in Tables~\ref{tab:qcID} and \ref{tab:eccID} after the
BHs had time to equilibrate ($t/m=200$).  
%Also provided are the difference in the masses,
%$\delta m = (q-1)/(q+1)$, sum of the spins $S/m^2 = (\chi_{2z} + q^2\chi_{1z})/(q+1)^2$, and spin difference, 
%$\Delta/m^2 = (\chi_{2z} - q\chi_{1z})/(1+q)$.
}
\label{tab:IDr}\\
\hline
\hline
Run & $q^r$ & $m^r_1/m$ & $m^r_2/m$ & $\chi^r_{1z}$ & $\chi^r_{2z}$  \\
\hline
\endfirsthead

\multicolumn{6}{c}
{{\tablename\ \thetable{} -- continued from previous page}}\\
\hline
Run & $q^r$ & $m^r_1/m$ & $m^r_2/m$ & $\chi^r_{1z}$ & $\chi^r_{2z}$  \\
\hline
\endhead
\hline \multicolumn{6}{r}{{Continued on next page}} \\
\hline
\endfoot
\hline\hline
\endlastfoot

RIT:BBH:0803 &	$0.2000$ & $0.1667$ & $0.8333$ & $0.2501$ & $0.3500$ \\
RIT:BBH:0804 &	$0.2000$ & $0.1666$ & $0.8333$ & $0.8006$ & $0.2500$ \\
RIT:BBH:0809 &	$0.2000$ & $0.1666$ & $0.8333$ & $-0.8006$ & $0.2500$ \\
RIT:BBH:0810 &	$0.2002$ & $0.1667$ & $0.8327$ & $-0.2501$ & $0.8514$ \\
RIT:BBH:0812 &	$0.0156$ & $0.0154$ & $0.9846$ & $-0.0000$ & $0.0000$ \\
RIT:BBH:0819 &	$0.2000$ & $0.1667$ & $0.8333$ & $-0.2501$ & $0.3500$ \\
RIT:BBH:0820 &	$0.9248$ & $0.4804$ & $0.5195$ & $-0.7003$ & $-0.4000$ \\
RIT:BBH:0825 &	$0.9248$ & $0.4804$ & $0.5195$ & $0.7003$ & $-0.4000$ \\
\end{longtable}

\clearpage
\begin{longtable*}{lcccrrrrrr}
\caption{The mass and spin of the 146 QC precessing BHBs in Table~\ref{tab:qcID} after the
BHs had time to equilibrate ($t/m=200$).}
\label{tab:IDr_prec}\\
\hline
\hline
Run & $q^r$ & $m^r_1/m$ & $m^r_2/m$ & $\chi^r_{1x}$ & $\chi^r_{1y}$ &  $\chi^r_{1z}$ &  $\chi^r_{2x}$ &  $\chi^r_{2y}$ & $\chi^r_{2z}$ \\
\hline
\endfirsthead

\multicolumn{10}{c}
{{\tablename\ \thetable{} -- continued from previous page}}\\
\hline
Run & $q^r$ & $m^r_1/m$ & $m^r_2/m$ & $\chi^r_{1x}$ & $\chi^r_{1y}$ &  $\chi^r_{1z}$ &  $\chi^r_{2x}$ &  $\chi^r_{2y}$ & $\chi^r_{2z}$ \\
\hline
\endhead
\hline \multicolumn{10}{r}{{Continued on next page}} \\
\hline
\endfoot
\hline\hline
\endlastfoot

RIT:BBH:0787 &	$0.5998$ & $0.3749$ & $0.6250$ & -0.3116 & 0.2257 & 0.7021 & - & - & - \\
RIT:BBH:0791 &	$0.0667$ & $0.0625$ & $0.9367$ & - & - & - & 0.6030 & 0.0334 & -0.6002 \\
RIT:BBH:0806 &	$0.4998$ & $0.3332$ & $0.6667$ & 0.3450 & 0.2137 & -0.6904 & - & - & - \\
RIT:BBH:0807 &	$0.2001$ & $0.1667$ & $0.8329$ & - & - & - & -0.3775 & -0.1337 & -0.6937 \\
RIT:BBH:0808 &	$0.2001$ & $0.1667$ & $0.8329$ & - & - & - & 0.1346 & -0.3819 & -0.6911 \\
RIT:BBH:0811 &	$0.5998$ & $0.3749$ & $0.6250$ & -0.3115 & -0.2618 & 0.6895 & - & - & - \\
RIT:BBH:0813 &	$0.2001$ & $0.1667$ & $0.8329$ & - & - & - & -0.6522 & -0.2397 & -0.3984 \\
RIT:BBH:0814 &	$0.5998$ & $0.3749$ & $0.6250$ & -0.3874 & 0.0232 & 0.7002 & - & - & - \\
\end{longtable*}

\clearpage
\begin{longtable}{lcr}
\caption{The energy radiated, $\delta \mathcal{M}^{IH} = M_{adm} - M_{rem}^{IH}$,
and final spin, $\chi_{\mathrm{rem}}^{IH}$, as measured using the IH formalism
for all simulations reported in this appendix. 
The error bars are due to variations in the measured mass and spin with time.  
For aligned systems with final spin antialigned to the initial orbital angular 
momentum the minus sign is preserved.}
\label{tab:spinerad}\\
\hline
\hline
Run & $\delta \mathcal{M}^{IH}$ & $\chi_{\mathrm{rem}}^{IH}$ \\
\hline
\hline
\endfirsthead

\multicolumn{3}{c}
{{\tablename\ \thetable{} -- continued from previous page}}\\
\hline
Run & $\delta \mathcal{M}^{IH}$ & $\chi_{\mathrm{rem}}^{IH}$ \\
\hline
\endhead
\hline \multicolumn{3}{r}{{Continued on next page}} \\
\hline
\endfoot
\hline\hline
\endlastfoot

RIT:BBH:0787 &  $0.048375 \pm 0.000001$ & 	$0.704948 \pm 0.000006$ \\
RIT:BBH:0791 &  $0.004042 \pm 0.000004$ & 	$-0.618003 \pm 0.000010$ \\
RIT:BBH:0803 &  $0.022730 \pm 0.000001$ & 	$0.624138 \pm 0.000013$ \\
RIT:BBH:0804 &  $0.021735 \pm 0.000003$ & 	$0.572977 \pm 0.000016$ \\
RIT:BBH:0806 &  $0.035674 \pm 0.000002$ & 	$0.582006 \pm 0.000038$ \\
RIT:BBH:0807 &  $0.012975 \pm 0.000025$ & 	$0.279147 \pm 0.000048$ \\
RIT:BBH:0808 &  $0.013010 \pm 0.000022$ & 	$0.278355 \pm 0.000037$ \\
RIT:BBH:0809 &  $0.019870 \pm 0.000001$ & 	$0.553156 \pm 0.000006$ \\
\end{longtable}

%%%%%%%%%%%%%%%%%%%%%%%%%%%%%%%%%%%%%%%%%%%%

\end{document}